\documentclass{aastex}
\usepackage{spr-astr-addons}
\usepackage{url}
\RequirePackage{color}

\newcommand{\emaila}{authors@email.com}

\begin{document}
\title{On estimators of the jet bolometric luminosity of Fermi 2LAC blazars}
\slugcomment{Not to appear in Nonlearned J., 45.}
%% Running heads
\shorttitle{Short article title}
\shortauthors{Wang et al.}
\author{Zerui Wang\altaffilmark{1}}\affil{Department of Physics, Yunnan Normal University, Kunming 650500, China}
\and \author{Rui Xue\altaffilmark{2}}
\affil{School of Astronomy and Space Science, Nanjing University, Nanjing 210093, China}
\and \author{Leiming Du\altaffilmark{1}$^{\dag}$}\and\author{Zhaohua Xie\altaffilmark{1}}\affil{Department of Physics, Yunnan Normal University, Kunming 650500, China}
\and \author{Dingrong Xiong\altaffilmark{3}}
\affil{National Astronomical Observatories/Yunnan Observatories, Chinese Academy of Sciences, Kunming 650011, China}
\and\author{Tingfeng Yi\altaffilmark{1}}\and\author{Yunbing Xu\altaffilmark{1}}\and\author{Wenguang Liu\altaffilmark{1}}
\affil{Department of Physics, Yunnan Normal University, Kunming 650500, China\\$^{\dag}$email:leiming\_du@ynao.ac.cn}
\email{\emaila}

\begin{abstract}
Bolometric luminosity is a basic physical parameter that is widely used in the study of blazars. Due to the lack of simultaneous full wavelength data, several estimators of the bolometric luminosity are being used in practice. In this paper, we study and evaluate the reliability and significance of six estimators, the $5GHz$ luminosity, the 1keV luminosity,  the $\gamma$-ray luminosity, the $5GHz$ luminosity+the 1keV luminosity, the $5GHz$ luminosity+the $\gamma$-ray luminosity and the 1keV luminosity+the $\gamma$-ray luminosity, by analyzing the linear correlations between the integrated bolometric luminosity and them. Our main results are as follows. (i) All the six estimators are reliable in the sense that they are all significant correlated with the bolometric luminosity. (ii) Ranking from the higher significance of the reliability to lower one the six estimators are the the $5GHz$ luminosity+the $\gamma$-ray luminosity, the 1keV luminosity+the $\gamma$-ray luminosity, $\gamma$-ray luminosity, the $5GHz$ luminosity+the 1keV luminosity, the $5GHz$ luminosity and the 1keV luminosity. (iii) We suggest that the bolometric luminosity can be well estimated by the $\gamma$-ray luminosity using the best linear equation that given in this paper for Fermi FSRQs. (iv) According to the linear regressions obtained in the analysis, we provide calibration for each estimator.
\end{abstract}
\section{Introduction}
The bolometric luminosity is the amount of electromagnetic energy a body radiates per unit of time, it is the total radiant energy over all wavelengths. Observationally, the typical broadband spectral energy distribution(SED) of a blazar displays a two-component structure in the log$\nu$-log$\nu F_{\nu}$(from radio band to $\gamma$ band ) diagram. In the SEDs of blazars, the non-thermal radiation from jet dominants at all wavelengths, and the thermal radiation from accretion onto the central black hole and from the host galaxy are prominent in the UV waveband(Marscher 2009; Abdo et al. 2010; Giommi et al. 2012; Ackermann et al. 2015). For blazars, bolometric luminosity usually refers to the non-thermal radiation from jet.

The shape of the SEDs dependes on the physical properties and structure of blazar jet. Bolometric luminosity is one of the most important basic parameters for blazars. It can be used to study the blazar sequence(Fossati et al. 1998, Ghisellini et al. 2008; Ghisellini et al. 2010; Ghisellini 2016; Mao et al. 2016; Fan et al. 2017) , the evolution of blazars(Xie et al. 2004; Xie et al. 2006), the jet-black hole relation and the jet-disc relation(Xiong et al. 2014; Yu et al. 2015; Xue et al. 2016).

A detailed SED can use observations from 10 or more telescopes, sometimes use more than one instrument on a single telescope(e.g., the Swift Gamma-Ray-Burst Explorer). In practice, because it is hard to get simultaneous data from full-wave band, many authors used estimation methods to calculate the bolometric luminosity.

The estimation methods that we study in this paper are as follows.

1. Fossati et al.(1998) study the SEDs of a sample of blazars using data from the radio to the $\gamma$-ray band. They find that as the bolometric luminosity increases, both peaks shift to lower frequencies. In their sample, part of blazars do not have the full-wave band data. Therefore, they use the 5GHz radio luminosity instead of the integrated bolometric luminosity which is available for all objects.

In Donato et al.(2001), the observed radio luminosity $L_{R}=(\nu L_{\nu})_{5GHz}$ is assumed to be linear proportional to the bolometric luminosity. In Beckmann et al.(2002) and Costamante $\&$ Ghisellini (2002), they also mention that the observed bolometric luminosity is thought to be well traced by the radio luminosity.

The Large Area Telescope (LAT) on board the Fermi Gamma Ray Space Telescope, launched on 2008 June 11, provides unprecedented sensitivity in the $\gamma$-ray band(20 MeV to over 300 GeV; Atwood et al. 2009). Information of the $\gamma$-ray band in the SED is available for more and more sources. However, possibility exists that there are still more blazars with no accurate data coverage for the IC component. At this time, information of the synchrotron component is important to estimate the jet bolometric luminosity. Therefore there is a need to find out whether the 5GHz radio luminosity can estimate the bolometric observed luminosity well for Fermi blazars.

2. One of the most important results of the CGRO/EGRET is the discovery that blazars emit most of their bolometric luminosity in the high gamma-rays(E$>$100MeV) energy range. Fossati et al.(1998) suggest that the $\gamma$-ray luminosity is closer to the integrated bolometric luminosity than the radio one. In Fan et al.(1999) and Xie et al. (2004), they study a sample of the EGRET detected blazars respectively. Because they considered the flare states of the selected objects, they take the gamma-ray luminosity to stand for half of the bolometric luminosity approximately, i.e. $L_{gamma}\sim0.5L_{bol}$.

Since the launch of the Fermi Large Area Telescope(LAT), more and more researchers use the observed gamma-ray luminosity as a proxy of the jet bolometric luminosity(Ghisellini et al. 2009, 2015, Nemmen et al. 2012, Sbarrato et al. 2012 and Pjanka et al.2017). When a blazar has no accurate data coverage for the synchrotron component or the IC component, the $\gamma$-ray luminosity will be an important estimator to calculate the bolometric luminosity for Fermi blazars. Therefore there is a need to find out whether the $\gamma$-ray luminosity can be a good estimator to indicate the bolometric luminosity.

3. In the study of AGNs, it is very common to use monochromatic luminosity to estimate the bolometric luminosity. In addition to 5GHz radio luminosity, it is meaningful to study whether other monochromatic luminosities(e.g. 5100$\mathring{A}$ optical luminosity and 1 keV X-ray luminosity) can estimate the jet bolometric luminosity well. For AGNs, bolometric luminosity is commonly estimated by scaling from a monochromatic luminosity at 5100$\mathring{A}$(Elvis et al. 1994; Kaspi et al. 2000; Shang et al. 2005; Richards et al. 2006; Trippe 2015). However, for blazars, 5100$\mathring{A}$ wavelength is very close to UV band which means that the luminosity at 5100$\mathring{A}$ will be contaminated by the thermal radiation. If one only get the data points around the UV band, one cannot know whether it is the non-thermal radiation from jet or the thermal radiation from disc/host galaxy. When studying the blazar jet, the method of using optical monochromatic luminosity to estimate the jet bolometric luminosity should be used very carefully. Therefore, in this paper, we study the linear correlation between the 1keV X-ray luminosity and the integrated bolometric luminosity. Meanwhile we also study whether it is better to use two luminosities  of 5GHz radio luminosity, the 1keV X-ray luminosity, the gamma-ray luminosity to estimate the integrated bolometric luminosity.

In this paper, the SEDs of both the synchrotron and IC components of a sample are fitted by a log-parabolic law. We measure the integrated bolometric luminosity by fitting the SEDs. We then present detailed studies on the correlations between the 5GHz radio luminosity, the 1keV luminosity, the $\gamma$-ray luminosity, the $5GHz$ luminosity+the 1keV luminosity, the $5GHz$ luminosity+the $\gamma$-ray luminosity, the 1keV luminosity+the $\gamma$-ray luminosity and the integrated bolometric luminosity. This paper is structured as follows. In Section 2, we present the sample. In Section 3, we present the results. In Section 4 we provide a discussion and conclusion.

\section{Sample}
Xue et al.(2016) collect a sample which contains 200 flat-spectrum radio quasars(FSRQs) and 79 BL Lacs from the second LAT AGN catalog(2LAC). At least one of the two components (synchrotron and IC components) of the blazars in their sample can be fitted by sufficient multifrequency data coverage. They construct the SEDs of all the blazars from the simultaneous multi-frequency data by using the ASDC SED Builder(It is an on-line service developed at the ASI Science Data Center. The database of the ASDC SED Builder provides observational data from many space telescopes and ground-based telescopes; Stratta et al. 2011). They use the second-degree polynomial function:
\begin{equation}
log(\nu F_{\nu})=c(log \nu)^2+b(log \nu)+a
\end{equation}
to fit the synchrotron component and IC component separately. The luminosities and peak frequencies (in the rest frame) can be calculated through $(\nu L_{\nu})_{syn,IC}=4{\pi}D_{L}^{2}(\nu F_{\nu})_{syn,IC}$ and ${\nu}_{syn,IC}=(1+z){\nu}_{syn,IC}^{obs}$, where $D_{L}$ is the luminosity distance and z is the redshift. In this paper we are going to study the estimation methods of the bolometric luminosity, therefore the 81 FSRQs and 28 BL Lacs with the complete SEDs in Xue et al.(2016) are taken as our sample.

According to the definition of bolometric luminosity, the SEDs can be numerically integrated over the entire frequency range to calculate the bolometric flux:
\begin{equation}
F_{integrated}=\int F_{\nu}d{\nu}=(ln10)\int 10^{(cx^2+bx+a)} dx.
\end{equation}
The bolometric luminosity can be calculated through:
\begin{equation}
L_{bol}=4\pi D_{L}^2 F_{integrated}.
\end{equation}

Detailed information about the sample is given in Table 1, with the following headings: column (1) name of the Fermi catalogue; (2) redshift; column (3) the observation date of the (quasi-) simultaneous data; (4) logarithm of the 5GHz radio luminosity in units of $erg s^{-1}$; (5) logarithm of the 1keV X-ray luminosity in units of $erg s^{-1}$; (6) logarithm of the $\gamma$-ray luminosity above 100 MeV in units of $erg s^{-1}$; (7) logarithm of the integrated bolometric luminosity in units of $erg s^{-1}$; (8) type of blazar.

The 5GHz radio luminosity, the 1keV X-ray luminosity and the $\gamma$-ray luminosity are got from the SEDs. For BL Lacs without a measured redshift, we assume the mean redshift value of 0.27 in 2LAC(Ackermann et al. 2011).

\section{Results of correlation analysis}
Linear regression and Pearson correlation analysis are applied to the relevant data to analyze the correlations between the 5GHz radio luminosity $log L_{5GHz}$, the 1keV X-ray luminosity $log L_{1keV}$, the $\gamma$-ray luminosity $log L_{\gamma}$ and the integrated bolometric luminosity $log L_{bol}$. The corresponding figures are given in Figs. 1, 2, 3. The analysis results and the best linear fitting equations(the black solid lines in Figs. 1, 2, 3) are as follows.

(1) $log L_{5GHz}$ versus $log L_{bol}$: $p<0.0001$, R=0.85, $log L_{bol}$=(0.80$\pm$0.04)$log L_{5GHz}$+(12.20$\pm$1.89).

(2) $log L_{1keV}$ versus $log L_{bol}$: $p<0.0001$, R=0.686, $log L_{bol}$=(1.14$\pm$0.08)$log L_{1keV}$+(-3.28$\pm$3.73).

(3) $log L_{\gamma}$ versus $log L_{bol}$: $p<0.0001$, R=0.975, $log L_{bol}$=(0.93$\pm$0.02)$log L_{\gamma}$+(3.68$\pm$0.95).

Meanwhile we use the multivariate regression method of Kelly(2007), which is also used in Plotkin et al. (2012) to get relationships between the $log L_{bol}$ and both the $log L_{5GHz}$ and $log L_{1keV}$, $log L_{5GHz}$ and $log L_{\gamma}$, $log L_{1keV}$ and $log L_{\gamma}$. Kelly (2007) use a Bayesian approach towards linear regression, estimating the probability distribution of the parameters with given observations. As has been concluded by Plotkin et al. (2012), the method of Kelly (2007) has advantage in accounting for correlated errors and capability providing statistical information for not just unknown parameters but also the intrinsic scatter. Moreover, the method of Kelly (2007) is suitable for handling heterogeneously selected data sets with large measurement uncertainties, and the measurement errors in the independent variables do not need to be of similar magnitude. The analysis results and the best linear fitting equations are as follows:

(1) $log L_{5GHz}$, $log L_{1keV}$ versus $log L_{bol}$: $p<0.0001$, R = 0.88, $log L_{bol}$ = (0.53 $\pm$ 0.06)$log L_{5GHz}$ + (0.52 $\pm$ 0.10)$log L_{1keV}$ + (0.82 $\pm$ 4.03).

(2) $log L_{5GHz}$, $log L_{\gamma}$ versus $log L_{bol}$: $p<0.0001$, R = 0.98, $log L_{bol}$ = (0.12 $\pm$ 0.09)$log L_{5GHz}$ + (0.82 $\pm$ 0.11)$log L_{\gamma}$ + (3.47 $\pm$ 3.03).

(3) $log L_{1keV}$, $log L_{\gamma}$ versus $log L_{bol}$: $p<0.0001$, R = 0.98, $log L_{bol}$ = (0.21 $\pm$ 0.11)$log L_{1keV}$ + (0.81 $\pm$ 0.08)$log L_{\gamma}$ + (-0.19 $\pm$ 3.71).

From the results we can see that all these six correlations are significant. Therefore, the integrated bolometric luminosity $L_{bol}$ can be estimated from $L_{5GHz}$, $log L_{1keV}$, $L_{\gamma}$, $log L_{5GHz}$ + $log L_{1keV}$, $log L_{5GHz}$ + $log L_{\gamma}$ and $log L_{1keV}$ + $log L_{\gamma}$ by using the best linear equations respectively.

\section{Discussion and conclusion}
By comparing the integrated bolometric luminosity with these estimation methods respectively, we can give a sequence(from the best to the worst): $log L_{5GHz}$ + $log L_{\gamma}$, $log L_{1keV}$ + $log L_{\gamma}$, the $\gamma$-ray luminosity, $log L_{5GHz}$ + $log L_{1keV}$, the 5GHz radio luminosity and the 1keV X-ray luminosity. Through studying the correlations between the 5GHz radio luminosity, the 1keV X-ray luminosity, the $\gamma$-ray luminosity, $log L_{5GHz}$ + $log L_{1keV}$, $log L_{5GHz}$ + $log L_{\gamma}$, $log L_{1keV}$ + $log L_{\gamma}$ and the integrated bolometric luminosity, we find that all these six correlations are significant for our blazar sample. Moreover we can find that the correlation coefficients of  $log L_{5GHz}$ + $log L_{\gamma}$, $log L_{1keV}$ + $log L_{\gamma}$ and the $\gamma$-ray luminosity are all very close to 0.98, which suggest that we can estimate the bolometric luminosity simply through using the $\gamma$-ray luminosity.

From our results, it can be found that the correlation between the $\gamma$-ray luminosity and the integrated bolometric luminosity is stronger than that between $log L_{5GHz}$, $log L_{1keV}$ and $log L_{bol}$. This means that even if only $\gamma$-ray data are available, $L_{\gamma}$ can serve as good estimator for the bolometric luminosity for Fermi blazars. In Fan et al. (1999) and Xie et al. (2004), they study the flare states of the selected objects, therefore they take the $\gamma$-ray luminosity to stand for half of the bolometric luminosity. From our result, it can be found that it is not accurate to estimate the bolometric luminosity with the twice the $\gamma$-ray luminosity. We suggest that using the best linear fitting equation is a better way to estimate the bolometric luminosity. The $\gamma$-ray band is one part of the IC component and $\gamma$-ray luminosity dominates the IC luminosity. Therefore, if the SED is dominated by the IC component, it is reasonable to use $\gamma$-ray luminosity to indicate the bolometric luminosity. By calculating the Compton dominance($CD=log(L_{IC}/L_{syn})$), it can be found that 77 FSRQs and 9 BL Lacs are dominated by the IC component in our sample. This means that our sample is dominated by FSRQs, and SEDs of our sample are dominated by the IC component. Perhaps the correlation between $\gamma$-ray luminosity and the bolometric luminosity will be influenced by selection effect, and maybe the correlation for FSRQs is different from that for BL Lacs. Therefore we also study this correlation for FSRQs and BL Lacs in our sample, respectively. We find that both of the correlations are still highly significant(FSRQs: R=0.97, $p<0.0001$; BL Lacs: 0.96, $p<0.0001$), and their best linear equation $log L_{bol}$=(0.93$\pm$0.03)$log L_{\gamma}$+(3.58$\pm$1.26) for FSRQs,  $log L_{bol}$=(0.93$\pm$0.05)$log L_{\gamma}$+(3.92$\pm$0.96) for BL Lacs are almost the same to the best linear equation of the whole sample. Correlation between $log L_{\gamma}$ and $log L_{bol}$ for BL Lacs is shown in Figure 4. In Figure 4 we can find that most of the data points in the elliptical region are the BL Lacs that their SEDs are dominated by the synchrotron component and they seem to show a trend that is completely different from the best linear fitting equation(the black solid line).  Therefore, we suggest that a larger sample of BL Lacs dominated by the synchrotron component is needed to check whether the correlation between $\gamma$-ray luminosity and the bolometric luminosity is still the same. In summary, we suggest that the bolometric luminosity can be well estimated by the $\gamma$-ray luminosity using the best linear equation given in this paper for FSRQs. 

In addition to these estimation methods mentioned above, Massaro et al. (2004) give another way to calculate the bolometric flux. The SED fitted by a log-parabolic law can be analytically integrated over the entire frequency range to calculate the bolometric flux. The final result to calculate the bolometric flux is (Massaro et al. 2004):
\begin{equation}
F_{bol}=\sqrt{\pi ln10}\frac{\nu_{p}F(\nu_{p})}{\sqrt{c}}=2.70\frac{\nu_{p}F(\nu_{p})}{\sqrt{c}},
\end{equation}
where $F_{bol}$ is the bolometric flux, $\nu_{p}F(\nu_{p})$ is the synchrotron/IC peak flux and c is the second-degree term of log-parabolic equation for the synchrotron/IC component.

One can find that this analytic expression is simple to calculate the bolometric luminosity. However, since Eq.(4) needs to know the curvature at the peak, which is a parameter that is not calculated and studied commonly. Therefore, few studies have used this method to estimate the bolometric luminosity before. Recently, since the curvature at the peak can be used to study the particle acceleration mechanisms, more and more researchers(Massaro et al.2004, 2006, Chen 2014, Xue et al. 2016, Fan et al. 2016) begin to calculate and study the curvature at the peak of the SED. Therefore, in the case of knowing the peak luminosity and curvature, we suggest to use Eq.(4) to calculate the bolometric luminosity instead of the integration method.

\begin{acknowledgements}
Part of this work is based on archival data, software or online services provided by the ASI SCIENCE DATA CENTER (ASDC). We are very grateful to Professor Giommi and the supporting team in ASDC for their help. This work is supported by the Joint Research Fund in Astronomy (Grant Nos U1431123, 10978019, 11263006, 11463001) under cooperative agreement between the National Natural Science Foundation of China (NSFC) and Chinese Academy of Sciences (CAS), the Provincial Natural Science Foundation of Yunnan (Grant No.: 2013FZ042) and the Yunnan province education department project (Grant No. 2014Y138, 2016ZZX066).
\end{acknowledgements}

\clearpage
\begin{figure}
\includegraphics[width=84mm]{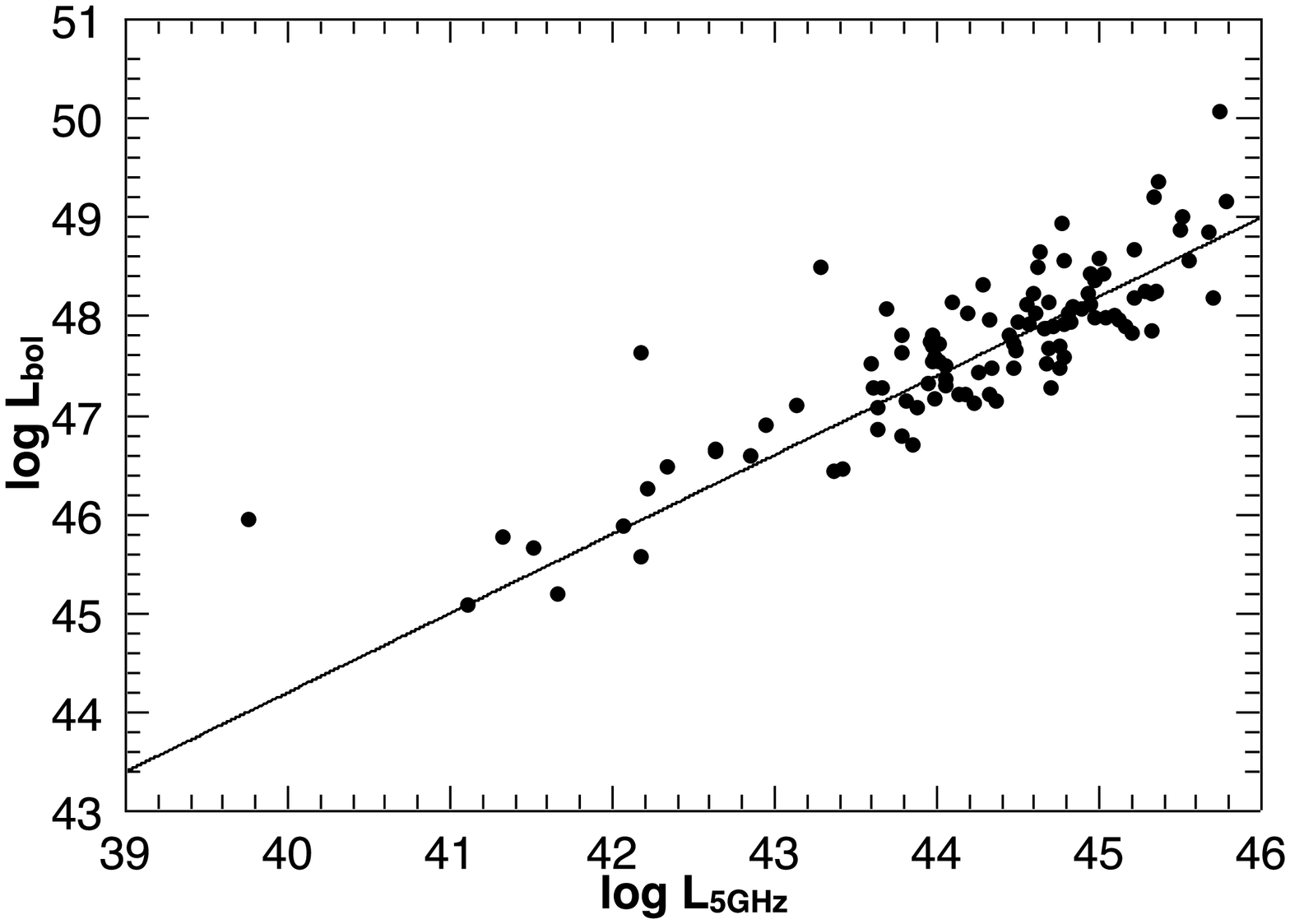}
\caption{Correlation between $log L_{5GHz}$ and $log L_{bol}$. The black solid line is result of linear regression.}
\end{figure}

\begin{figure}
\includegraphics[width=84mm]{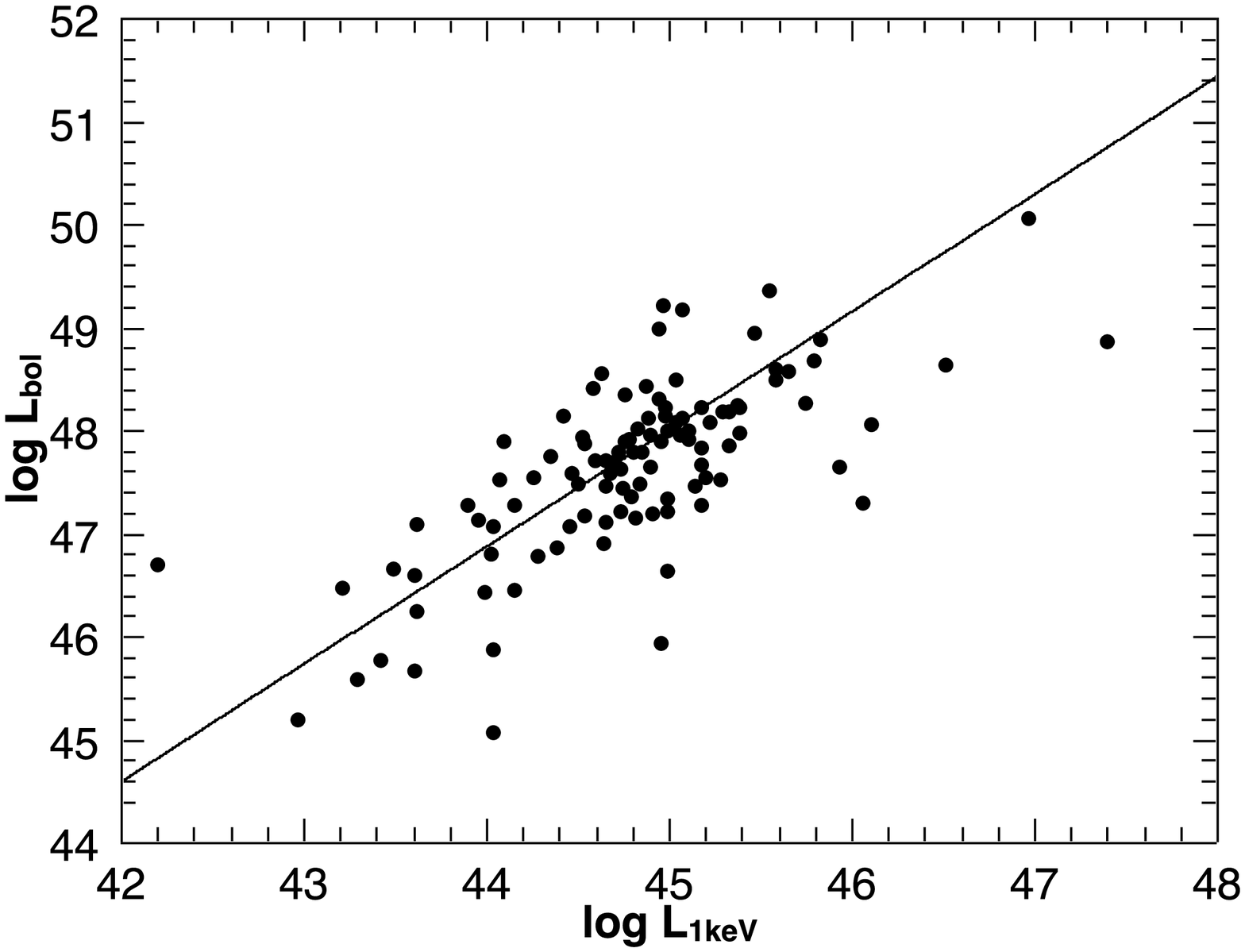}
\caption{Correlation between $log  L_{1keV}$ and $log L_{bol}$. The black solid line is result of linear regression.}
\end{figure}

\begin{figure}
\includegraphics[width=84mm]{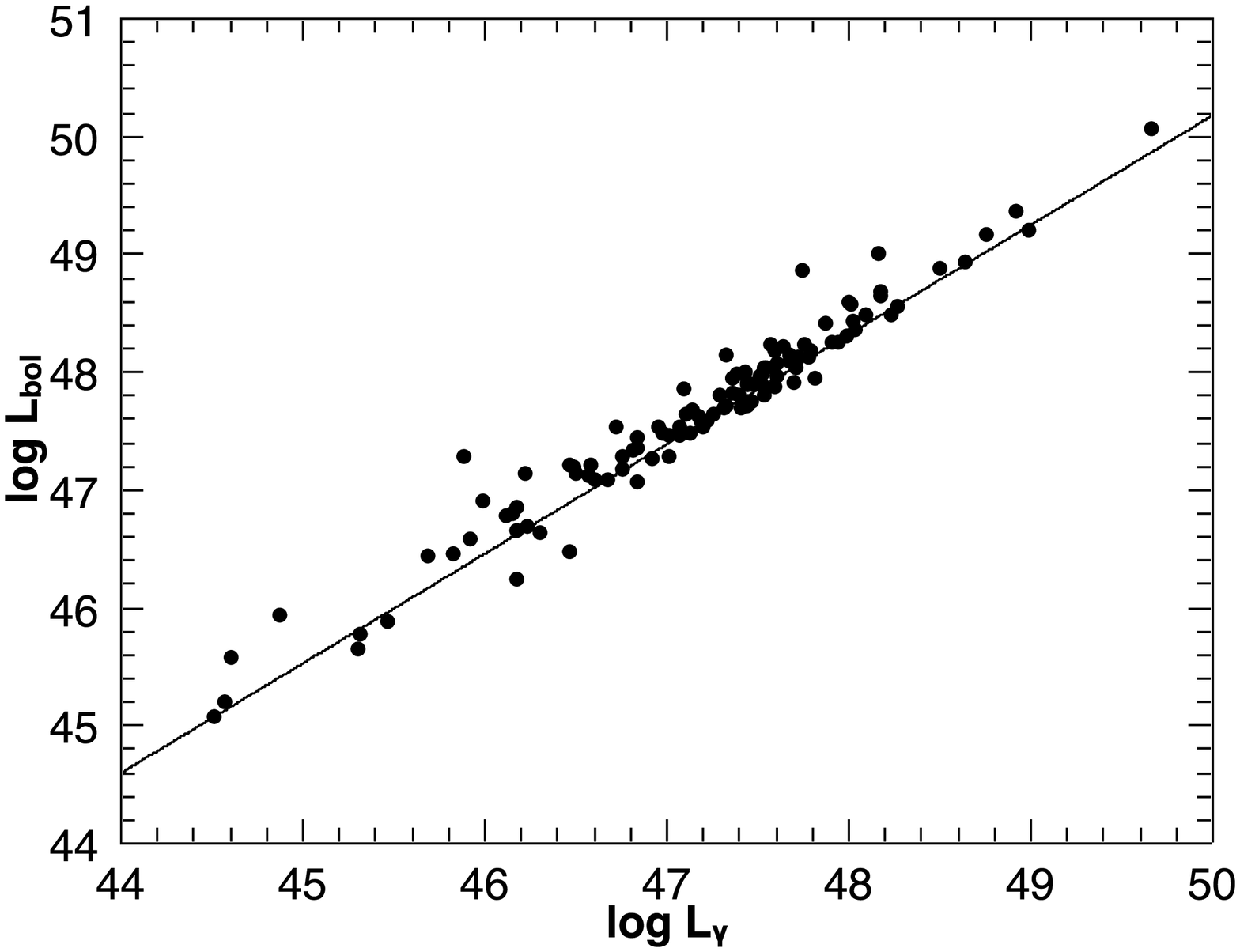}
\caption{Correlation between $log L_{\gamma}$ and $log L_{bol}$. The black solid line is result of linear regression.}
\end{figure}

\begin{figure}
\includegraphics[width=84mm]{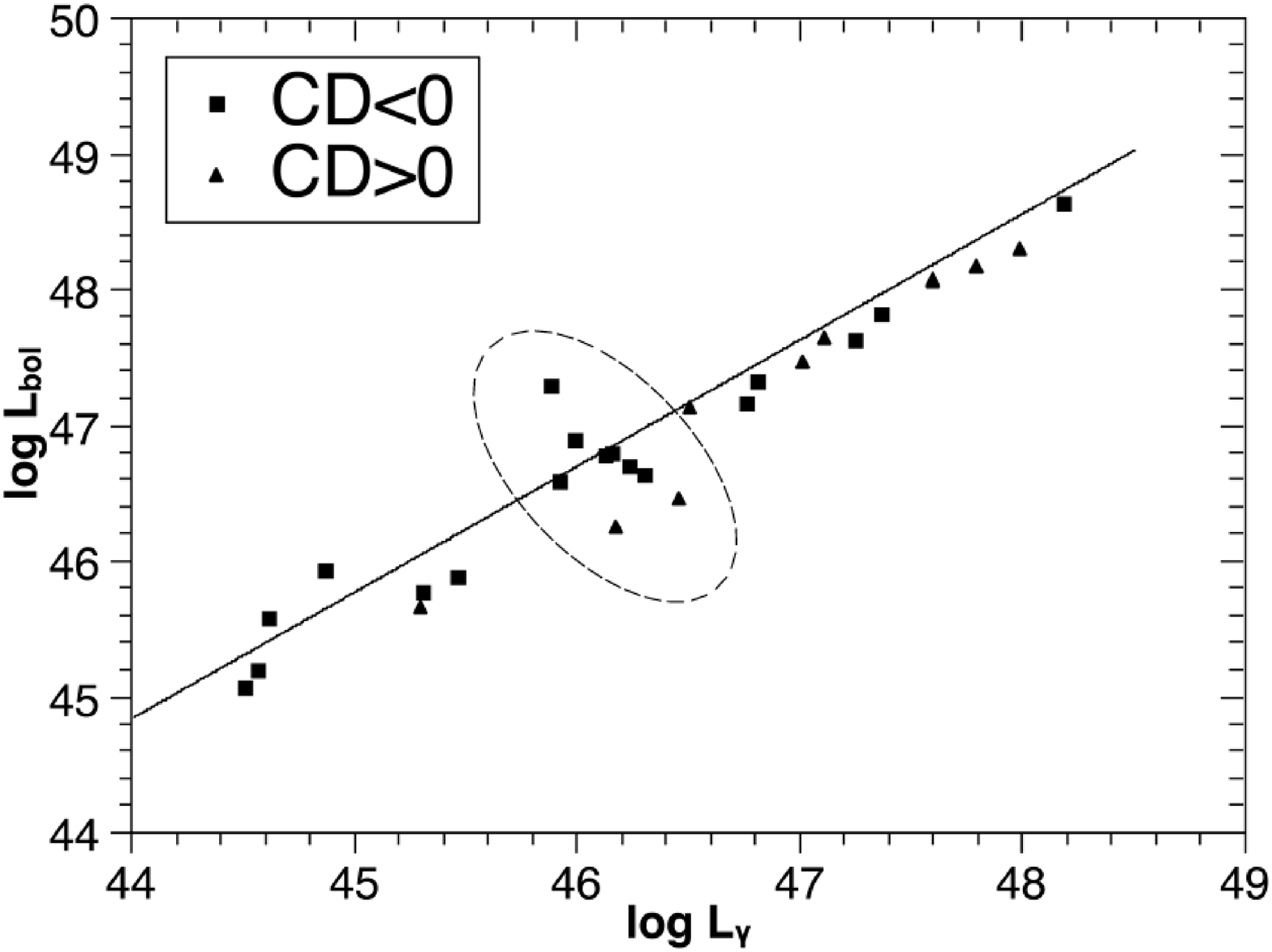}
\caption{Correlation between $log L_{\gamma}$ and $log L_{bol}$ for BL Lacs. The black solid line is result of linear regression.  }
\end{figure}

\clearpage
\begin{deluxetable}{crccccccccccccrl}
\tabletypesize{\scriptsize} \rotate \tablecaption{The sample of 109 Fermi blazars\label{tbl-1}} \tablewidth{0pt}
\tablehead{\colhead{Fermi Name} & \colhead{Redshift} & \colhead{time} & \colhead{$LogL_{5GHz}$} & \colhead{$LogL_{1keV}$}  & \colhead{$LogL_{\gamma}$} & \colhead{${LogL_{bol}}$} & \colhead{blazar type}\\
\colhead{(1)} & \colhead{(2)} & \colhead{(3)} & \colhead{(4)} &
\colhead{(5)} & \colhead{(6)} & \colhead{(7)} & \colhead{(8)} } \startdata
2FGL J0017.4-0018	&	1.574	&	2010/6/24	&	44.71 	&	44.76 	&	47.44 	&	47.90 	&	FSRQ	\\
2FGL J0049.7-5738	&	1.797	&	2010/5/26-2010/6/10	&	45.33 	&	45.33 	&	47.09 	&	47.85 	&	FSRQ	\\
2FGL J0051.0-0648	&	1.975	&	2009/5/17	&	45.28 	&	45.37 	&	47.91 	&	48.25 	&	FSRQ	\\
2FGL J0109.9+6132	&	0.785	&	2010/1/31-2010/2/3	&	44.10 	&	44.42 	&	47.67 	&	48.15 	&	FSRQ	\\
2FGL J0113.7+4948	&	0.395	&	2010/8/27	&	43.36 	&	43.99 	&	45.69 	&	46.44 	&	FSRQ	\\
2FGLJ0136.9+4751	&	0.859	&	2010/2/5	&	45.04 	&	44.99 	&	47.52 	&	47.99 	&	FSRQ	\\
2FGL J0158.0-4609	&	2.287	&	2010/5/2	&	44.66 	&	44.53 	&	47.59 	&	47.88 	&	FSRQ	\\
2FGL J0205.3-1657	&	1.466	&	2010/1/8	&	44.97 	&	45.38 	&	47.38 	&	47.98 	&	FSRQ	\\
2FGL J0217.7+7353	&	2.367	&	2010/9/10-2010/9/11	&	45.74 	&	46.97 	&	49.66 	&	50.07 	&	FSRQ	\\
2FGLJ0217.9+0143	&	1.715	&	2010/1/29-2010/2/2	&	45.35 	&	45.74 	&	47.94 	&	48.26 	&	FSRQ	\\
2FGL J0221.0+3555	&	0.944	&	2010/8/4-2010/8/20	&	44.02 	&	44.59 	&	47.44 	&	47.72 	&	FSRQ	\\
2FGL J0222.0-1615	&	0.7	&	2010/5/30	&	43.63 	&	44.45 	&	46.61 	&	47.08 	&	FSRQ	\\
2FGL J0237.1-6136	&	0.467	&	2010/6/12-2010/6/16	&	43.29 	&	45.58 	&	48.23 	&	48.49 	&	FSRQ	\\
2FGLJ0237.8+2846	&	1.213	&	2010/2/5	&	45.03 	&	44.58 	&	47.87 	&	48.42 	&	FSRQ	\\
2FGL J0245.1+2406	&	2.247	&	2010/02/01-2010/3/11	&	44.77 	&	45.47 	&	48.64 	&	48.94 	&	FSRQ	\\
2FGL J0407.7+0740	&	1.133	&	2010/2/14-2010/2/15	&	44.06 	&	44.50 	&	46.98 	&	47.49 	&	FSRQ	\\
2FGLJ0423.2-0120	&	0.916	&	2009/8/27	&	45.33 	&	45.18 	&	47.64 	&	48.22 	&	FSRQ	\\
2FGL J0439.0-1252	&	1.285	&	2010/7/1	&	44.34 	&	44.65 	&	47.07 	&	47.47 	&	FSRQ	\\
2FGLJ0457.0-2325	&	1.003	&	2010/2/25	&	44.71 	&	44.09 	&	47.52 	&	47.90 	&	FSRQ	\\
2FGL J0532.7+0733	&	1.254	&	2010/4/25	&	44.94 	&	45.07 	&	47.72 	&	48.12 	&	FSRQ	\\
2FGLJ0539.3-2841	&	3.104	&	2010/3/12	&	45.51 	&	44.94 	&	48.16 	&	49.00 	&	FSRQ	\\
2FGL J0601.1-7037	&	2.409	&	2010/3/9	&	44.79 	&	44.63 	&	48.27 	&	48.56 	&	FSRQ	\\
2FGL J0608.0-0836	&	0.872	&	2010/6/7	&	44.70 	&	45.18 	&	47.01 	&	47.28 	&	FSRQ	\\
2FGL J0654.2+4514	&	0.928	&	2010/3/23	&	43.98 	&	44.46 	&	47.22 	&	47.59 	&	FSRQ	\\
2FGL J0654.5+5043	&	1.253	&	2010/1/15	&	43.97 	&	44.69 	&	47.41 	&	47.69 	&	FSRQ	\\
2FGL J0656.2-0320	&	0.634	&	2010/3/31	&	43.88 	&	44.03 	&	46.84 	&	47.07 	&	FSRQ	\\
2FGL J0714.0+1933	&	0.54	&	2010/4/2-2010/4/3	&	43.13 	&	43.62 	&	46.67 	&	47.09 	&	FSRQ	\\
2FGL J0746.6+2549	&	2.979	&	2010/10/15	&	45.21 	&	45.79 	&	48.17 	&	48.68 	&	FSRQ	\\
2FGL J0750.6+1230	&	0.889	&	2010/4/1-2010/4/12	&	44.68 	&	45.28 	&	46.72 	&	47.53 	&	FSRQ	\\
2FGL J0805.5+6145	&	3.033	&	2010/4/3-2010/4/4	&	45.56 	&	45.65 	&	48.01 	&	48.57 	&	FSRQ	\\
2FGL J0909.1+0121	&	1.024	&	2010/4/18-2010/5/5	&	44.60 	&	45.38 	&	47.57 	&	48.23 	&	FSRQ	\\
2FGL J0912.1+4126	&	2.563	&	2010/2/25	&	44.33 	&	44.89 	&	47.60 	&	47.96 	&	FSRQ	\\
2FGL J0920.9+4441	&	2.19	&	2009/10/29	&	45.50 	&	45.82 	&	48.50 	&	48.88 	&	FSRQ	\\
2FGL J0956.9+2516	&	0.707	&	2010/5/7-2010/6/15	&	44.18 	&	44.73 	&	46.58 	&	47.22 	&	FSRQ	\\
2FGL J0957.7+5522	&	0.896	&	2009/11/1	&	44.50 	&	44.52 	&	47.81 	&	47.94 	&	FSRQ	\\
2FGL J1037.5-2820	&	1.066	&	2010/1/22-2010/1/23	&	44.01 	&	44.25 	&	47.07 	&	47.54 	&	FSRQ	\\
2FGL J1106.1+2814	&	0.843	&	2010/5/24	&	43.66 	&	44.15 	&	46.76 	&	47.28 	&	FSRQ	\\
2FGLJ1126.6-1856	&	1.048	&	2010/6/10	&	44.79 	&	44.78 	&	47.44 	&	47.92 	&	FSRQ	\\
2FGL J1146.8-3812	&	1.048	&	2010/6/24	&	44.69 	&	45.17 	&	47.14 	&	47.68 	&	FSRQ	\\
2FGL J1159.5+2914	&	0.724	&	2010/5/28-2010/6/11	&	44.47 	&	44.65 	&	47.32 	&	47.72 	&	FSRQ	\\
2FGL J1206.0-2638	&	0.789	&	2010/6/30-2010/12/10	&	44.14 	&	44.91 	&	46.49 	&	47.20 	&	FSRQ	\\
2FGL J1208.8+5441	&	1.344	&	2010/5/16-2010/5/21	&	44.19 	&	44.83 	&	47.53 	&	48.03 	&	FSRQ	\\
2FGLJ1222.4+0413	&	0.967	&	2010/6/17-2010/6/29	&	44.69 	&	44.98 	&	47.32 	&	48.14 	&	FSRQ	\\
2FGLJ1246.7-2546	&	0.635	&	2010/1/25	&	43.79 	&	44.73 	&	47.18 	&	47.63 	&	FSRQ	\\
2FGLJ1256.1-0547	&	0.536	&	2010/1/15	&	45.12 	&	45.06 	&	47.51 	&	47.96 	&	FSRQ	\\
2FGLJ1310.6+3222	&	0.997	&	2009/12/12-2009/12/21	&	45.16 	&	44.95 	&	47.48 	&	47.89 	&	FSRQ	\\
2FGL J1326.8+2210	&	1.4	&	2010/3/30	&	44.81 	&	45.02 	&	47.55 	&	48.03 	&	FSRQ	\\
2FGL J1333.5+5058	&	1.362	&	2010/3/13	&	43.59 	&	44.07 	&	47.20 	&	47.53 	&	FSRQ	\\
2FGL J1337.7-1257	&	0.539	&	2010/1/18-2010/1/26	&	44.33 	&	44.99 	&	46.47 	&	47.22 	&	FSRQ	\\
2FGL J1347.7-3752	&	1.3	&	2010/9/20	&	43.97 	&	44.80 	&	47.29 	&	47.80 	&	FSRQ	\\
2FGL J1358.1+7644	&	1.585	&	2010/5/24-2010/5/25	&	44.75 	&	44.70 	&	47.31 	&	47.70 	&	FSRQ	\\
2FGL J1408.8-0751	&	1.494	&	2010/5/23	&	44.84 	&	45.04 	&	47.67 	&	48.09 	&	FSRQ	\\
2FGL J1436.9+2319	&	1.548	&	2010/6/14	&	44.78 	&	44.67 	&	47.19 	&	47.59 	&	FSRQ	\\
2FGL J1504.3+1029	&	1.839	&	2010/7/29	&	45.34 	&	44.96 	&	48.99 	&	49.21 	&	FSRQ	\\
2FGL J1514.6+4449	&	0.57	&	2010/4/6	&	42.63 	&	43.49 	&	46.17 	&	46.66 	&	FSRQ	\\
2FGL J1539.5+2747	&	2.191	&	2010/3/17	&	44.57 	&	45.11 	&	47.70 	&	47.92 	&	FSRQ	\\
2FGL J1549.5+0237	&	0.414	&	2010/2/13-2010/2/20	&	43.63 	&	44.38 	&	46.18 	&	46.86 	&	FSRQ	\\
2FGLJ1635.2+3810	&	1.814	&	2010/3/7	&	45.79 	&	45.07 	&	48.75 	&	49.17 	&	FSRQ	\\
2FGL J1637.7+4714	&	0.735	&	2010/7/30-2010/7/31	&	44.06 	&	44.79 	&	46.84 	&	47.36 	&	FSRQ	\\
2FGLJ1640.7+3945	&	1.66	&	2010/8/7	&	44.97 	&	44.75 	&	48.03 	&	48.36 	&	FSRQ	\\
2FGL J1709.7+4319	&	1.027	&	2009/12/1	&	43.78 	&	44.85 	&	47.40 	&	47.80 	&	FSRQ	\\
2FGL J1733.1-1307	&	0.902	&	2010/3/14-2010/4/10	&	45.09 	&	45.11 	&	47.43 	&	48.01 	&	FSRQ	\\
2FGL J1848.5+3216	&	0.798	&	2010/10/6-2010/10/19	&	43.97 	&	45.20 	&	46.95 	&	47.54 	&	FSRQ	\\
2FGLJ1911.1-2005	&	1.119	&	2009/10/4	&	44.93 	&	44.98 	&	47.75 	&	48.23 	&	FSRQ	\\
2FGL J1924.8-2912	&	0.353	&	2010/9/30	&	44.36 	&	44.81 	&	46.22 	&	47.15 	&	FSRQ	\\
2FGL J1958.2-3848	&	0.63	&	2010/4/9-2010/4/14	&	44.26 	&	44.74 	&	46.84 	&	47.44 	&	FSRQ	\\
2FGL J1959.1-4245	&	2.178	&	2010/4/5-2010/4/14	&	44.62 	&	45.04 	&	48.09 	&	48.49 	&	FSRQ	\\
2FGL J2135.6-4959	&	2.181	&	2010/4/22-2010/5/5	&	44.55 	&	44.88 	&	47.78 	&	48.12 	&	FSRQ	\\
2FGL J2144.8-3356	&	1.361	&	2009/9/22-2009/9/24	&	43.96 	&	44.35 	&	47.47 	&	47.75 	&	FSRQ	\\
2FGLJ2151.5-3021	&	2.345	&	2010/5/4-2010/5/13	&	45.67 	&	47.40 	&	47.74 	&	48.86 	&	FSRQ	\\
2FGL J2157.4+3129	&	1.488	&	2009/7/8-2009/7/12	&	44.61 	&	44.83 	&	47.71 	&	48.03 	&	FSRQ	\\
2FGL J2201.9-8335	&	1.865	&	2010/7/5-2010/7/17	&	44.94 	&	44.87 	&	48.02 	&	48.43 	&	FSRQ	\\
2FGL J2211.9+2355	&	1.125	&	2009/4/15-2009/4/21	&	44.47 	&	44.84 	&	47.13 	&	47.48 	&	FSRQ	\\
2FGL J2225.6-0454	&	1.404	&	2010/5/22-2010/5/27	&	45.70 	&	45.32 	&	47.59 	&	48.19 	&	FSRQ	\\
2FGLJ2253.9+1609	&	0.859	&	2009/12/4-2009/12/6	&	45.37 	&	45.55 	&	48.92 	&	49.36 	&	FSRQ	\\
2FGL J2258.0-2759	&	0.926	&	2010/5/20-2010/5/26	&	44.82 	&	44.89 	&	47.36 	&	47.95 	&	FSRQ	\\
2FGL J2322.2+3206	&	1.489	&	2009/5/20	&	44.45 	&	44.72 	&	47.53 	&	47.80 	&	FSRQ	\\
2FGL J2327.5+0940	&	1.841	&	2010/6/18-2010/6/29	&	45.00 	&	45.58 	&	48.00 	&	48.59 	&	FSRQ	\\
2FGL J2334.3+0734	&	0.401	&	2009/12/20	&	43.42 	&	44.15 	&	45.83 	&	46.45 	&	FSRQ	\\
2FGLJ2345.0-1553	&	0.621	&	2009/1/10	&	43.61 	&	43.89 	&	46.92 	&	47.27 	&	FSRQ	\\
2FGL J2347.9-1629	&	0.576	&	2009/12/04-2009/12/05	&	44.23 	&	44.65 	&	46.57 	&	47.12 	&	FSRQ	\\
2FGL J0050.6-0929	&	0.635	&	2009/5/24	&	44.05 	&	46.06	&	45.88 	&	47.29 	&	BL Lac	\\
2FGL J0100.2+0746	&	0.27	&	2010/7/9	&	42.34 	&	43.21	&	46.46 	&	46.47 	&	BL Lac	\\
2FGLJ0114.7+1326 	&	0.27	&	2010/7/15	&	43.69 	&	46.10	&	47.60 	&	48.07 	&	BL Lac	\\
2FGL J0141.5-0928	&	0.733	&	2010/5/30-2010/6/6	&	43.95 	&	44.99 	&	46.81 	&	47.33 	&	BL Lac	\\
2FGL J0153.9+0823	&	0.27	&	2010/3/1	&	42.63 	&	44.99	&	46.30 	&	46.64 	&	BL Lac	\\
2FGLJ0210.7-5102	&	1.003	&	2009/11/26	&	44.89 	&	45.22 	&	47.60 	&	48.08 	&	BL Lac	\\
2FGLJ0238.7+1637	&	0.94	&	2010/1/30	&	44.29 	&	44.94 	&	47.99 	&	48.31 	&	BL Lac	\\
2FGLJ0334.2-4008	&	1.445	&	2010/1/17-2010/1/18	&	45.22 	&	45.29 	&	47.79 	&	48.18 	&	BL Lac	\\
2FGLJ0523.0-3628	&	0.055	&	2010/3/5	&	42.17 	&	43.29 	&	44.61 	&	45.58 	&	BL Lac	\\
2FGLJ0712.9+5032 	&	0.27	&	2009/1/21	&	42.85 	&	43.61 	&	45.92 	&	46.59 	&	BL Lac	\\
2FGLJ0721.9+7120	&	0.27	&	2005/4/4	&	44.63 	&	46.51	&	48.18 	&	48.64 	&	BL Lac	\\
2FGLJ0854.8+2005	&	0.306	&	2010/4/10	&	43.78 	&	44.28	&	46.12 	&	46.78 	&	BL Lac	\\
2FGLJ0958.6+6533	&	0.367	&	2010/3/12	&	42.95 	&	44.64 	&	45.99 	&	46.90 	&	BL Lac	\\
2FGLJ1043.1+2404	&	0.559117	&	2010/7/9	&	43.79 	&	44.02 	&	46.15 	&	46.80 	&	BL Lac	\\
2FGLJ1058.4+0133	&	0.888	&	2009/12/3	&	45.20 	&	45.18 	&	47.36 	&	47.83 	&	BL Lac	\\
2FGLJ1104.4+3812	&	0.03	&	2009/11/15-2009/11/17	&	39.76 	&	44.95	&	44.87 	&	45.94 	&	BL Lac	\\
2FGLJ1217.8+3006	&	0.13	&	2009/12/3-2009/12/19	&	42.07 	&	44.03	&	45.46 	&	45.88 	&	BL Lac	\\
2FGLJ1146.8-3812	&	1.048	&	2010/6/24	&	44.76 	&	45.14 	&	47.01 	&	47.47 	&	BL Lac	\\
2FGLJ1221.4+2814	&	0.102	&	2009/12/10-2009/12/12	&	41.33 	&	43.42	&	45.31 	&	45.78 	&	BL Lac	\\
2FGLJ1248.2+5820	&	0.8474	&	2010/5/20	&	42.18 	&	45.93	&	47.25 	&	47.64 	&	BL Lac	\\
2FGLJ1517.7-2421	&	0.048	&	2010/2/20	&	41.66 	&	42.96 	&	44.57 	&	45.20 	&	BL Lac	\\
2FGLJ1542.9+6129	&	0.117	&	2009/1/18-2009/1/20	&	41.52 	&	43.60	&	45.30 	&	45.66 	&	BL Lac	\\
2FGLJ1653.9+3945	&	0.033	&	2010/3/21	&	41.11 	&	44.03	&	44.51 	&	45.08 	&	BL Lac	\\
2FGLJ1719.3+1744	&	0.137	&	2009/1/8	&	42.21 	&	43.62 	&	46.17 	&	46.25 	&	BL Lac	\\
2FGLJ1751.5+0938	&	0.322	&	2010/4/1	&	43.85 	&	42.20 	&	46.23 	&	46.70 	&	BL Lac	\\
2FGLJ1800.5+7829	&	0.68	&	2009/10/13	&	44.48 	&	44.90 	&	47.11 	&	47.65 	&	BL Lac	\\
2FGLJ2152.4+1735	&	0.871	&	2010/4/8-2010/4/10	&	43.81 	&	43.95 	&	46.50 	&	47.14 	&	BL Lac	\\
2FGLJ2247.2-0002	&	0.949	&	2010/1/14-2010/1/16	&	43.98 	&	44.54 	&	46.76 	&	47.17 	&	BL Lac	\\
\enddata
\end{deluxetable}
\end{document}